\documentclass{article}

\usepackage[utf8]{inputenc}
\usepackage[T1]{fontenc}
\usepackage{geometry}
\usepackage{booktabs}
\usepackage{multirow}
\usepackage{hyperref}
\usepackage{amsmath}
\usepackage{xcolor}
\usepackage[numbers]{natbib}
\usepackage{caption}
\usepackage{graphicx} % Required for inserting images
\usepackage{authblk}
\usepackage{enumitem}
\setlist[itemize]{itemsep=0.2ex, topsep=0.5ex}
\setlist[enumerate]{itemsep=0.2ex, topsep=0.5ex}

\hypersetup{
    colorlinks=true,
    linkcolor=blue!60!black,
    citecolor=blue!60!black,
    urlcolor=blue!60!black
}

\title{Evaluation of Music Version Retrieval Methods \\for Setlist Identification in Live Concerts}

\author{Andres Ferraro, Albin Correya}
\affil{Nowon AG, Switzerland}
\affil{\texttt{research@nowon.io}}

\date{August 2026}

\begin{document}

\maketitle

% ============================================================
\section{Introduction}
% ============================================================
Setlist identification is the task of automatically determining which songs are performed during a live concert recording by matching segments of the concert audio against a database of known reference tracks.
This is a specific application of \emph{music version identification} (VI), where the goal is to recognize a song despite the significant acoustic differences between a query song and its original version.

Nowon focuses on research related to version identification and copyright infringement detection in AI-generated music. The production use case at Nowon imposes challenges that go beyond standard VI benchmarks:
\begin{itemize}
    \item \textbf{Low audio quality.} Concert recordings range from professional recordings to smartphone captures in noisy venues, introducing reverberation, crowd noise, clipping, and low-frequency rumble.
    \item \textbf{Arrangement variations.} Live performances often differ substantially from studio versions through key changes, tempo variations, extended solos, mashups, medleys, and shortened or rearranged sections.
    \item \textbf{Scale.} A single concert can last 1--3 hours, and the
reference database may contain millions of candidate songs, requiring
efficient retrieval.
    \item \textbf{Continuous monitoring.} Unlike query-by-example retrieval where a clean query is provided to identify a single song, setlist identification must process the full concert timeline producing detections over time.
\end{itemize}

This report provides an objective evaluation of three version identification systems on the ASID benchmark~\cite{yesiler2021}, comparing their effectiveness under these conditions.

% ============================================================
\section{Related Work}
% ============================================================

\subsection{Music Version Identification}
Music version identification has been studied extensively in MIR.
Early approaches used hand-crafted chroma features with sequence
alignment (e.g., DTW, cross-similarity)~\cite{ellis2007,serra2008,bertin2012}.
Deep embedding models then enabled scalable nearest-neighbor retrieval.
MOVE~\cite{yesiler2020move} and Re-MOVE~\cite{yesiler2020} learn
compact embeddings from cremaPCP chroma; CQTNet~\cite{yu2020},
ByteCover~\cite{du2021bytecover,du2022bytecover2,du2023bytecover3},
and CoverHunter~\cite{liu2023coverhunter} explore alternative
architectures and losses on standard cover-song benchmarks such as
SHS100K~\cite{xu2018} and Da-TACOS~\cite{yesiler2019datacos}.
More recently, CLEWS~\cite{serra2025} uses contrastive learning on
segment-level embeddings with distance-reduction strategies
(e.g., \texttt{bpwr}) for variable-length matching.

\subsection{Setlist Identification}
The specific problem of setlist identification --- matching concert audio against references over
a continuous timeline---has received less attention.
An early study by \textbf{Wang et al.}~\cite{wang2014} explicitly formulated the task as automatic setlist identification and song segmentation for full-length concert videos. Their approach sequentially identifies songs from a database of studio recordings and estimates their boundaries, and was evaluated on a dataset of 20 full concerts and 1,152 studio tracks.

More recently, \textbf{Yesiler et al.}~\cite{yesiler2021}, revisit the task and introduce the \textbf{ASID} (Automatic Setlist IDentification) dataset and benchmark.

The method from \textbf{Yesiler et al.}~\cite{yesiler2021} uses a sliding window over the concert audio and matches
each window against the full reference database.
Temporal postprocessing then merges consecutive detections of the same
song and resolves overlaps by keeping the highest-confidence match.
Optionally, an SVM trained on the development set filters remaining
false positives.
\textbf{Yesiler et al.}~\cite{yesiler2021} evaluates Qmax~\cite{serra2008}, 2DFTM~\cite{bertin2012}, and
Re-MOVE~\cite{yesiler2020}, with Re-MOVE obtaining the best performance.

To our knowledge, ASID is the largest publicly available benchmarks specifically designed for setlist
identification under the noisy, variable-quality conditions typical of
live concerts.

% ============================================================
\section{Evaluation Setup}
% ============================================================
\subsection{Dataset: ASID}
The ASID dataset was not available at the time of conducting this work. Therefore, we had to collect the original audio and regenerate the features to reproduce the results. However, some of the original concerts were not available anymore.

The successfully reconstructed ASID dataset~\cite{yesiler2021} used contains: 
\begin{itemize}
    \item \textbf{75 concerts}\footnote{The original ASID dataset contains 75 concerts, comprising 10 development and 65 evaluation concerts. At the time of our reproduction, three evaluation concerts were no longer recoverable, resulting in a reconstructed evaluation subset of 62 concerts} (10 development) spanning ${\sim}100$ hours of audio across 5 genres: pop (14), rock/metal (21), indie/alternative (15), hip-hop/rap (8), and electronic (7).
    \item \textbf{1,298 reference songs} (studio recordings) associated with the performing artists.
    \item \textbf{Three audio quality tiers:} AQ-A (broadcast/official, 32 concerts), AQ-B (professional small-venue recordings, 18 concerts), and AQ-C (smartphone/camera captures, 15 concerts).
    \item Ground-truth annotations mapping each song occurrence in a concert to its reference, with start and end timestamps.
\end{itemize}
\subsection{Evaluation Protocol}
Following the ASID benchmark protocol:
\begin{enumerate}
    \item A sliding window of \textbf{120\,s} with a \textbf{30\,s} hop is applied over each concert.
    \item Each window is scored against all reference songs; the top-scoring reference is selected.
    \item Temporal postprocessing merges consecutive windows matching the same reference, resolves overlapping detections by keeping the highest-confidence match, and merges again.
    \item Results are evaluated with two metrics.
\end{enumerate}
\subsection{Metrics}
\begin{itemize}
    \item \textbf{DAP (Detection Accuracy Percentage):}
    The fraction of ground-truth song annotations that are detected (i.e., have at least one overlapping true-positive match):
    \[
        \text{DAP} = \frac{\text{annotations} - \text{FN}}{\text{annotations}}
    \]
    DAP measures recall at the annotation level. False positives do not affect DAP directly.
    \item \textbf{DLP (Detected Length Percentage):}
    The fraction of total annotated duration covered by temporally overlapping true-positive detections:
    \[
        \text{DLP} = \frac{\sum \text{overlap}(\text{TP}_i, \text{GT}_j)}{\sum \text{duration}(\text{GT}_j)}
    \]
    DLP captures how well detections cover the actual song boundaries.
\end{itemize}

We also report window-level true positives (TP), false positives (FP), and false negatives (FN) after postprocessing.

\subsection{SVM Classifier}

The ASID paper includes an optional second stage: an SVM trained on development-set TP/FP features (match duration, confidence) to filter false positives on the eval set.

We report results both with and without this classifier to separate the contribution of the core VI system from the post-hoc filtering.

% ============================================================
\subsection{Compared Methods}
Although several systems above outperform Re-MOVE on track-level
cover-song benchmarks, they are rarely validated on long, noisy
concerts.
We therefore restrict this study to three systems that are most relevant
to setlist identification:
(i)~\textbf{Re-MOVE}~\cite{yesiler2020}, the strongest baseline in the
original ASID benchmark~\cite{yesiler2021} and the closest published prior
work on live-concert matching;
(ii)~\textbf{CLEWS}~\cite{serra2025}, a recent state-of-the-art VI
embedding model; and
(iii)~\textbf{NAVE}, our internal system.
Our goal is not an exhaustive comparison of all VI systems, but a fair comparison under the
ASID protocol against the established  baseline and a current
SOTA embedding method.
\paragraph{Re-MOVE (Baseline):}
Re-MOVE~\cite{yesiler2020} is the strongest system in the original ASID
paper.
It operates on cremaPCP chroma features (12-dimensional), produces
fixed-dimensional embeddings via a metric-learning model, and scores
query--reference pairs using cosine distance.
We use the original authors' code and pre-trained model.

\paragraph{CLEWS:}

CLEWS~\cite{serra2025} uses a contrastive learning framework producing segment-level embeddings.
Scoring uses normalized squared Euclidean distance between all query--reference shingle pairs, reduced via bpwr-5 or bpwr-10.

We use the publicly released checkpoint from Zenodo with the official \texttt{clews} inference code.
CLEWS operates on raw audio (spectrogram internally) and produces multiple embeddings of 1024 dimensions that represent each song.

\paragraph{NAVE (Nowon Internal):}
% \textcolor{gray}{fine-tuned from a CLEWS-family architecture}.
We evaluate NAVE (Neural Audio Version Embeddings) Nowon's internal system using two scoring modes:
\begin{itemize}
    \item \textbf{Recognition:} Production scoring pipeline using L2 distance with diagonal-band dynamic programming
    \item \textbf{Pooled:} Mean-pooled segment embeddings --- a Re-MOVE-like ablation that tests the embedding quality without the DP alignment.
\end{itemize}

Both modes use L2-normalized segment embeddings of 256 dimensions.

Table~\ref{tab:dims} summarizes the embedding dimensionalities.
NAVE matches Re-MOVE's compact 256-D representation while
retaining CLEWS-style segment-level matching.
\begin{table}[ht]
\centering
\caption{Embedding dimensionality of the compared systems.}
\label{tab:dims}
\begin{tabular}{lccc}
\toprule
\textbf{System} & \textbf{Input} & \textbf{Embedding} & \textbf{Dim.} \\
\midrule
Re-MOVE & cremaPCP chroma & track / window vector & 256 \\
CLEWS   & audio & segment sequence & 1024 \\
NAVE   & audio & segment sequence & 256 \\
\bottomrule
\end{tabular}
\end{table}
% ============================================================
\section{Results}
% ============================================================

\subsection{Overall Comparison}
Table~\ref{tab:overall} shows the main results on the ASID evaluation set (62 concerts), including the reproduced results for Re-MOVE and the ones provided in the original paper. All NAVE and CLEWS results do not apply SVM filtering.

Re-MOVE results are shown both with and without the SVM, as in the original paper.

\begin{table}[ht]
\centering
\caption{Overall results on ASID eval set (62 concerts, 120\,s / 30\,s sliding window, postprocessed). Re-MOVE numbers, both with/without SVM classifier variant. NAVE and CLEWS are evaluated without SVM.}
\label{tab:overall}
\begin{tabular}{llcccccc}
\toprule
\textbf{System} & \textbf{Filter} & \textbf{DAP} & \textbf{DLP} & \textbf{TP} & \textbf{FP} & \textbf{FN} \\
\midrule
Re-MOVE (paper)        & none       & 0.803 & 0.605 & --- & --- & --- \\
Re-MOVE (paper)        & SVM        & 0.678 & 0.569 & --- & --- & --- \\
Re-MOVE (ours)         & none       & 0.779 & 0.573 & 860 & 1053 & 240 \\
Re-MOVE (ours)         & SVM        & 0.601 & 0.507 & 652 & \textbf{115}  & 432 \\
\midrule
CLEWS             & none       & 0.903 & 0.689 & 975 & 272  & 109 \\
\midrule
NAVE recognition      & none       & 0.903 & 0.697 & 973 & 280  & 105 \\
NAVE pooled           & none       & \textbf{0.910} & \textbf{0.705} & \textbf{985} & 346  & \textbf{98} \\\bottomrule
\end{tabular}
\end{table}

\paragraph{Key observations.}
\begin{itemize}
    \item All CLEWS and NAVE systems achieve over ${\sim}0.90$ DAP, a $+12$ percentage point improvement compared to Re-MOVE ($0.78$--$0.80$ DAP).
    \item NAVE recognition achieves a higher DLP than CLEWS despite using a much lower dimensionality (256 vs 1024).
    \item NAVE pooled achieves higher DAP than NAVE recognition ($0.910$) but at the cost of more false positives ($346$ vs.\ $280$).
    \item The SVM classifier substantially reduces Re-MOVE's FP count ($1053 \to 115$) but also degrades DAP ($0.779 \to 0.601$), indicating aggressive filtering.
\end{itemize}

\subsection{Results by Audio Quality}
\begin{table}[ht]
\centering
\caption{Eval DAP by audio quality tier (NAVE recognition, no SVM).}
\label{tab:aq}
\begin{tabular}{lccccc}
\toprule
\textbf{Quality} & \textbf{DAP} & \textbf{TP} & \textbf{FP} & \textbf{FN} & \textbf{$n$ concerts} \\
\midrule
AQ-A (broadcast)   & 0.897 & 505 & 165 & 58 & 31 \\
AQ-B (professional rec.)   & 0.971 & 199 & 47 & 6 & 16 \\
AQ-C (smartphone)  & 0.862 & 254 & 63 & 41 & 14 \\
\bottomrule
\end{tabular}
\end{table}
\paragraph{Key observation.}
AQ-B concerts yield the highest DAP ($0.971$), while AQ-C smartphone recordings are the most challenging ($0.862$).
Surprisingly, AQ-A broadcast recordings ($0.897$) are not the easiest tier --- this may reflect that broadcast mixes include more crowd noise, commentary, and audio processing than clean pro-venue recordings.

\subsection{Results by Genre}

Table~\ref{tab:genre} compares all three systems across ASID genre categories on the eval set.
\begin{table}[ht]
\centering
\caption{Eval DAP and DLP by genre (postprocessed, no SVM).}
\label{tab:genre}
\begin{tabular}{lcccccc}
\toprule
& \multicolumn{2}{c}{\textbf{CLEWS}}
& \multicolumn{2}{c}{\textbf{NAVE recog.}}
& \multicolumn{2}{c}{\textbf{NAVE pooled}} \\
\cmidrule(lr){2-3} \cmidrule(lr){4-5} \cmidrule(lr){6-7}
\textbf{Genre} & DAP & DLP & DAP & DLP & DAP & DLP \\
\midrule
Pop           & 0.929 & 0.829 & \textbf{0.933} & \textbf{0.839} & 0.921 & 0.832 \\
Rock/Metal    & 0.886 & 0.799 & \textbf{0.903} & \textbf{0.816} & 0.897 & 0.811 \\
Indie/Alt.    & 0.943 & 0.796 & 0.943 & 0.822 & \textbf{0.953} & \textbf{0.830} \\
Hip-hop/Rap   & \textbf{0.845} & \textbf{0.278} & 0.810 & 0.273 & 0.796 & 0.271 \\
Electronic    & 0.891 & 0.686 & 0.873 & 0.691 & \textbf{0.982} & \textbf{0.791} \\
\bottomrule
\end{tabular}
\end{table}
\paragraph{Key observations.}
\begin{itemize}
    \item \textbf{Hip-hop/rap is the weakest category} for all systems (DAP $0.80$--$0.85$, DLP ${\sim}0.27$). This is expected since rap songs rely heavily on vocal content and rhythmic patterns rather than melodic/harmonic structure, which tonal embeddings capture less effectively.
    \item \textbf{Electronic} shows interesting divergence. We observe a high performance of NAVE pooled while NAVE recognition has a lower performance ($0.873$), suggesting the DP alignment may be less suited to repetitive electronic structures.
    \item \textbf{Indie/alternative} is the best category (${\sim}0.94$--$0.95$) with low false positives across all systems.
    \item System differences are small ($<5$ pp) on most categories.
\end{itemize}

\subsection{Music/Non-Music Gating}
We additionally filter NAVE recognition windows with a music/non-music
classifier before computing setlist metrics.
We use the TVSM-pseudo CRNN~\cite{hung2022}, a frame-level speech and music
activity detector trained on the TVSM dataset.
The goal is to suppress false positives in non-music segments (applause,
speech, silence).

We compare four gating strategies on top of the same CRNN predictions:
\begin{itemize}
    \item \texttt{overlap\_mean}: keep a window if at least 50\% of its
    duration overlaps the music mask (mean-fused CRNN chunks).
    \item \texttt{music\_beats\_speech}: same 50\% overlap gate, but the
    mask requires music to dominate speech.
    \item \texttt{conservative\_gate}: smoother mask (hysteresis, minimum
    music duration) and a stricter 80\% overlap requirement.
    \item \texttt{music\_island\_loose}: keep windows only inside contiguous
    ``music islands'' seeded by the CRNN (more aggressive clipping).
\end{itemize}

\begin{table}[ht]
\centering
\caption{Effect of TVSM music-gating strategies on NAVE recognition (eval, 62 concerts).}
\label{tab:gating}
\begin{tabular}{lccccc}
\toprule
\textbf{Strategy} & \textbf{DAP} & \textbf{DLP} & \textbf{TP} & \textbf{FP} & \textbf{FN} \\
\midrule
Unfiltered                    & 0.903 & 0.697 & 973 & 280 & 105 \\
\texttt{overlap\_mean}        & \textbf{0.909} & 0.705 & \textbf{984} & 331 & \textbf{99} \\
\texttt{music\_beats\_speech} & \textbf{0.909} & 0.705 & \textbf{984} & 313 & \textbf{99} \\
\texttt{conservative\_gate}   & 0.908 & \textbf{0.707} & 983 & 287 & 100 \\
\texttt{music\_island\_loose} & 0.839 & 0.640 & 924 & \textbf{243} & 177 \\
\bottomrule
\end{tabular}
\end{table}

\paragraph{Key observations.}
\begin{itemize}
    \item \texttt{conservative\_gate} offers the best trade-off and the highest DLP, with FP counts close to the unfiltered baseline.
    \item \texttt{music\_island\_loose} is too aggressive for 120\,s sliding
    windows, increasing number of FNs, reducing DAP by 6\,pp, and DLP by 5.7\,pp.
    \item Music gating provides a modest refinement rather than a substantial improvement, confirming that the VI system itself handles noisy segments reasonably well.
\end{itemize}

\section{Qualitative Analysis of Recognition Boundaries}

From the results presented in table \ref{tab:gating}, we see that most metrics are not much affected by applying the gating. However, after taking a closer look at some samples, we identified that there is a very important difference on the final recognition output.

\begin{figure}[h]
    \centering
    \includegraphics[width=0.95\textwidth]{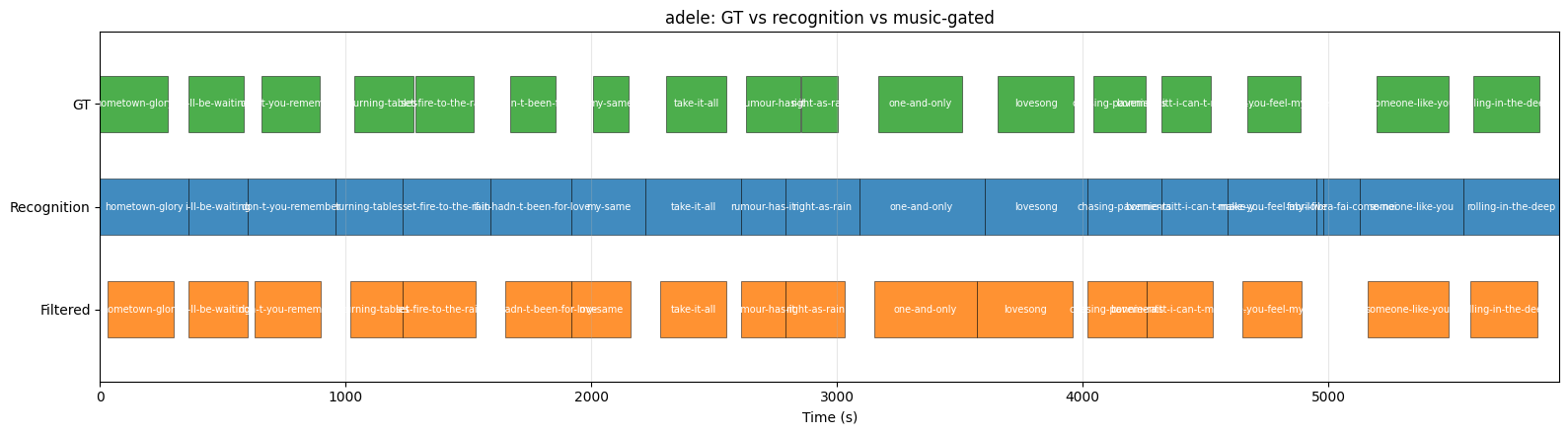}
    \caption{Full version identification output using NAVE recognition filtered vs unfiltered for Adele concert.}
    \label{fig:image1}
\end{figure}

\begin{figure}[h]
    \centering
    \includegraphics[width=0.95\textwidth]{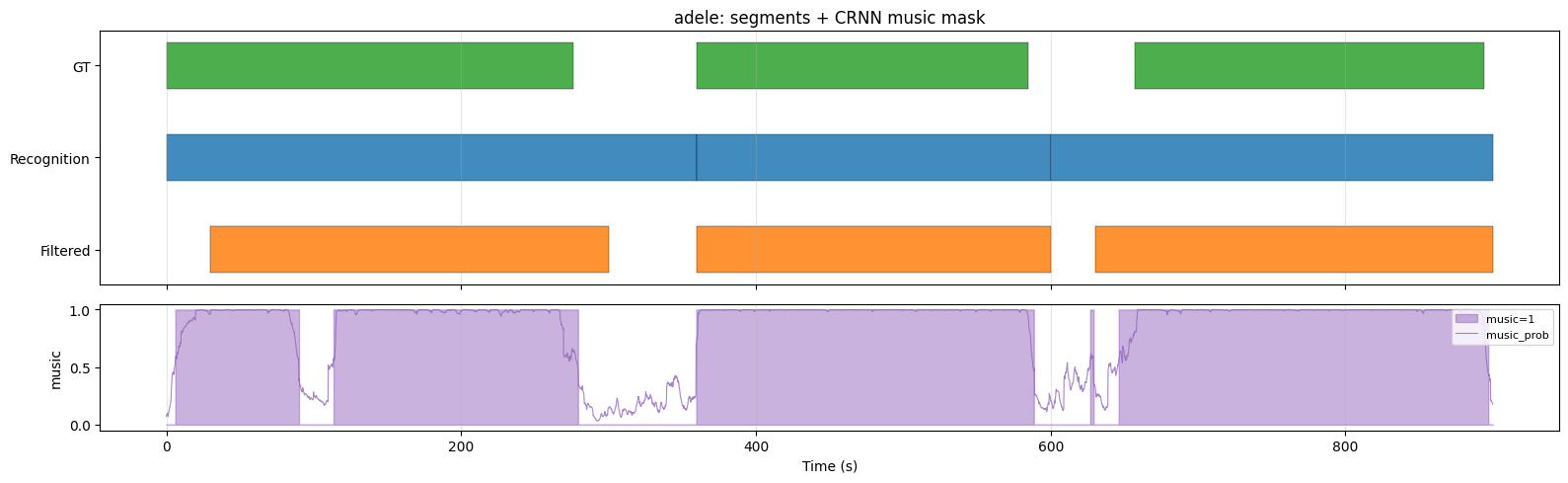}
    \caption{Version identification output using NAVE recognition filtered vs unfiltered, and score of the music/non-music classifier. Zoom in on first part of Adele concert.}
    \label{fig:image2}
\end{figure}

On Figures \ref{fig:image1} and \ref{fig:image2}, we can see that the final recognition is much more similar to the original ground truth annotations on the boundaries between songs. This raises a potential limitation on the evaluation protocol used in the previous ASID paper, and the need for a specific metric that allows to capture such difference.

% ============================================================
\section{Discussion}
% ============================================================
\subsection{CLEWS, NAVE, and Re-MOVE}
The most striking result is the ${\sim}12$ pp DAP improvement of all NAVE and CLEWS systems over Re-MOVE.
This gap likely stems from three factors:
\begin{enumerate}
    \item \textbf{Representation:} CLEWS and NAVE models operate on  Constant-Q Transform (CQT) spectrogram representations rather than hand-crafted cremaPCP chroma, capturing richer timbral and spectral information.
    \item \textbf{Shingle-level matching:} Rather than pooling an entire 120\,s window into a single embedding, CLEWS and NAVE produce sequences of shingle embeddings (${\sim}20$\,s each), enabling finer-grained temporal matching that is more robust to partial overlaps and arrangement changes.
    \item \textbf{Scalability:} NAVE embeddings are 256-dimensional, a $4\times$ reduction compared to CLEWS's 1024-dimensional shingles. This reduces storage and retrieval costs while slightly improving identification accuracy.
\end{enumerate}
\subsection{NAVE Recognition vs.\ Pooled}
The recognition scorer (DP alignment) and the pooled scorer (mean + cosine) achieve similar DAP but differ in precision:
\begin{itemize}
    \item Recognition: 280 FP vs.\ Pooled: 346 FP (eval), a 24\% difference.
    \item The DP alignment penalizes matches that lack temporal consistency, acting as an implicit false-positive filter.
    \item Pooled achieves slightly higher DAP ($+0.7$ pp) by being more permissive, catching a few extra annotations at the cost of noise.
\end{itemize}

\subsection{Impact of Audio Quality}
The results confirm that audio quality significantly affects identification accuracy.
AQ-C (smartphone) recordings are the hardest ($-11$ pp vs.\ AQ-B), consistent with the noise, reverberation, and spectral distortion typical of crowd-recorded audio.
This validates that our evaluation covers realistic deployment conditions.

\subsection{Storage and Serving Efficiency}
In production systems, embedding size matters for indexing large catalogs. Here we provide a comparison of the storage efficiency between the different systems.

Assuming float32 segment embeddings with a 20\,s window and 5\,s hop for  both CLEWS and NAVE pipelines, a 5-minute reference yields
57 shingles.

Under that setup, storing one million songs requires approximately
\textbf{233\,GB} for CLEWS (1024-D) versus \textbf{58\,GB} for NAVE
(256-D)---a $4\times$ reduction with no loss in DAP and a slight gain
in DLP (Table~\ref{tab:overall}).
By comparison, Re-MOVE stores a single 256-D vector per track
(${\sim}$1\,GB for the same catalog), but that compactness comes with
the ${\sim}$12\,pp DAP gap observed on ASID.

In production, the NAVE footprint also reduces memory for
nearest-neighbor search and lowers I/O when streaming embeddings for
long concerts, while retaining segment-level matching that Re-MOVE's
single-vector representation does not provide.

\subsection{Limitations and Future Work}
\begin{itemize}
    \item \textbf{Dataset size:} ASID contains 75 concerts with ${\sim}1{,}300$ references. While this is the largest public setlist-ID benchmark, it may not capture the full diversity of production conditions.
    \item \textbf{Missing concerts:} 3 eval concerts lacked audio at evaluation time, reducing the effective eval set to 62 concerts.
    \item \textbf{No SVM for CLEWS/NAVE:} To isolate the effect of the VI system, CLEWS and NAVE are evaluated using the same sliding-window and temporal postprocessing protocol. We do not train the original ASID SVM for these systems because its features and score distributions were designed around Re-MOVE.
    \item \textbf{Reference coverage:} Some reference songs were unavailable or had extraction failures, potentially affecting recall for specific concerts.
    \item \textbf{Evaluation methods:} After looking closer to the output of the recognition systems, we see that the metrics used in the original ASID paper may not fully capture the temporal precision of the recognition between the boundaries.
    \item \textbf{Metric saturation:} NAVE system already reach
    ${\sim}0.90$ DAP on ASID. At this level, small DAP differences seem hard to interpret and may reflect dataset limitations, more than clear model gains. We may be approaching a ceiling of the current ASID dataset/metrics.
\end{itemize}
% ============================================================
\section{Conclusion}
% ============================================================
This evaluation demonstrates that NAVE embedding models substantially outperform Re-MOVE on setlist identification, achieving ${\sim}0.90$ DAP vs.\ ${\sim}0.78$ on the ASID benchmark.
Nowon's NAVE recognition model with 256 dimensions improves the public CLEWS checkpoint in detection accuracy while being substantially smaller than CLEWS's 1024-dimensional representation.

Music/non-music gating provides a modest additional improvement, and could be integrated as a lightweight preprocessing step in production. Through qualitative analysis, we identified improvement from music/non-music gating as well as some potential limitations of the metrics used in the original ASID paper. Additionally, the preprocessing step provides an advantage for real-world long concert recordings, as it can reduce computational costs by avoiding expensive processing of short and long gaps between music performances. Therefore, efficiently identifying non-music sections can provide additional benefits when processing long audio streams, such as recordings lasting two hours or longer. 

% Mention limitations here and higlight that we try to do our best to make it public and objective experiments
% What is the message for the public audience
% Maybe we reach a limit of the metrics or the dataset, what does it mean to get 0.9 DAP?
% Higlight that the no one has done this comparison with CLEWS
% Higlight that the advantage compared to CLEWS is a reduced size, make it more efficient. Make this clear on the table so is easy to see
% mention that there are parts that are not here compared, because in production we have other components
% 

%============================================================
\bibliographystyle{unsrtnat}

\end{document}